\documentclass[prb,twocolumn,superscriptaddress,showpacs]{revtex4}
\usepackage{amsmath}
\usepackage{bm}
\usepackage{graphicx}
\usepackage{amssymb}
\usepackage{color}
\begin{document}
\newcommand{\figwidth}{0.95\columnwidth}
\newcommand{\ffigwidth}{0.4\columnwidth}
\newcommand{\warwick}{Department of Physics and Centre for Scientific Computing, University of Warwick, Coventry,
  CV4 7AL, United Kingdom}
\newcommand{\usal}{Departamento de Fisica Fundamental, Universidad de Salamanca, 37008 Salamanca, Spain}
\title{Multifractal analysis of the metal-insulator transition in the 3D Anderson model II: Symmetry relation under ensemble averaging}
\author{Alberto Rodriguez}
\affiliation{\warwick}
\affiliation{\usal}
\author{Louella J. Vasquez}
\affiliation{\warwick}
\author{Rudolf A. R\"omer}
\email[Corresponding author: ]{R.Roemer@warwick.ac.uk}
\affiliation{\warwick}
%
%
\begin{abstract}
We study the multifractal analysis (MFA) of electronic wavefunctions at the localisation-delocalisation transition in the 3D Anderson model for very large system sizes up to $240^3$. The singularity spectrum $f(\alpha)$ is numerically obtained using the \textsl{ensemble average} of the scaling law for the generalized inverse participation ratios $P_q$, employing box-size and system-size scaling. The validity of a recently reported symmetry law  [Phys. Rev. Lett. 97, 046803 (2006)] for the multifractal spectrum  is carefully analysed at the metal-insulator transition (MIT). The results are compared to those obtained using different approaches, in particular the typical average of the scaling law. System-size scaling with ensemble average appears as the most adequate method to carry out the numerical MFA. Some conjectures about the true shape of $f(\alpha)$ in the thermodynamic limit are also made.
\end{abstract}
\maketitle
\section{Introduction}
\label{sec-intro}
The multifractal analysis (MFA) of electronic wavefunctions $\vert\psi_i\vert^2$ at the localisation-delocalisation transition has been a subject of intense numerical study for the last twenty years. \cite{Aok83,SouE84,Sch85,Aok86,CasP86,OnoOK89,BauCS90,SchG91,GruS93,Jan94a,GruS95,MilRS97,MilEM02, SubGLE06,MilE07}
The MFA of the critical  $|\psi_i|^2$ in a system with volume $L^d$ is based on the scaling of the generalized inverse participation
ratios (gIPR), $P_q (l)\equiv\sum_{k}\mu_k^q(l)$, defined from the integrated measure $\mu_k(l)=\sum_i |\psi_i|^2$ in all $N_l$  boxes with linear size $l$ covering the system. At criticality the scaling law  $P_q (\lambda)\propto\lambda^{\tau(q)}$  is expected to hold in a certain range of values for  $\lambda\equiv l/L$.
The well-known singularity spectrum $f(\alpha)$ is defined from the $\tau(q)$ exponents via a Legendre transformation, $f(\alpha_q)=q\alpha_q-\tau(q)$ and $\alpha_q=\tau^\prime(q)$.  The physical meaning of the $f(\alpha)$ is as follows.  It is the fractal dimension of the set of points where the wavefunction intensity obeys
$|\psi_i|^2\sim L^{-\alpha}$, that is in a discrete system the number $N_\alpha$ of such points scales as $L^{f(\alpha)}$.

Recently, the report of a remarkable analytical result concerning the existence of an exact symmetry relation in the $f(\alpha)$,\cite{MirFME06} has required a profound revision of all the techniques involved in the numerical MFA.
The reported symmetry law\cite{MirFME06} requires $\Delta_q=\Delta_{1-q}$ in terms of the anomalous scaling exponents, which can be obtained by $\Delta_q=\tau(q) -d(q-1)$.
The symmetry can also be written as
\begin{equation}
	\alpha_q+\alpha_{1-q} =2d,
	\label{eq-symalfaq}
\end{equation}
or in terms of the singularity spectrum itself as
\begin{equation}
	\delta f(\alpha)\equiv \left\vert f(2d-\alpha)-[f(\alpha)+d-\alpha]\right\vert=0.
	\label{eq-symfalfa}
\end{equation}
At the metal-insulator transition (MIT) the $f(\alpha)$ is a convex function of $\alpha\geqslant 0$, with a  maximum at $\alpha_0 \geqslant d$ where $f(\alpha_0)=d$. The values
of $f(\alpha)$ are not restricted to be positive. \cite{ChhS91,Man03} The symmetry \eqref{eq-symfalfa} implies that the upper bound is $\alpha\leqslant 2d$ and that the values of $f(\alpha)$ for $\alpha<d$ can be
mapped to the values for $\alpha>d$, and vice versa.
\begin{figure}
 	\includegraphics[width=\figwidth]{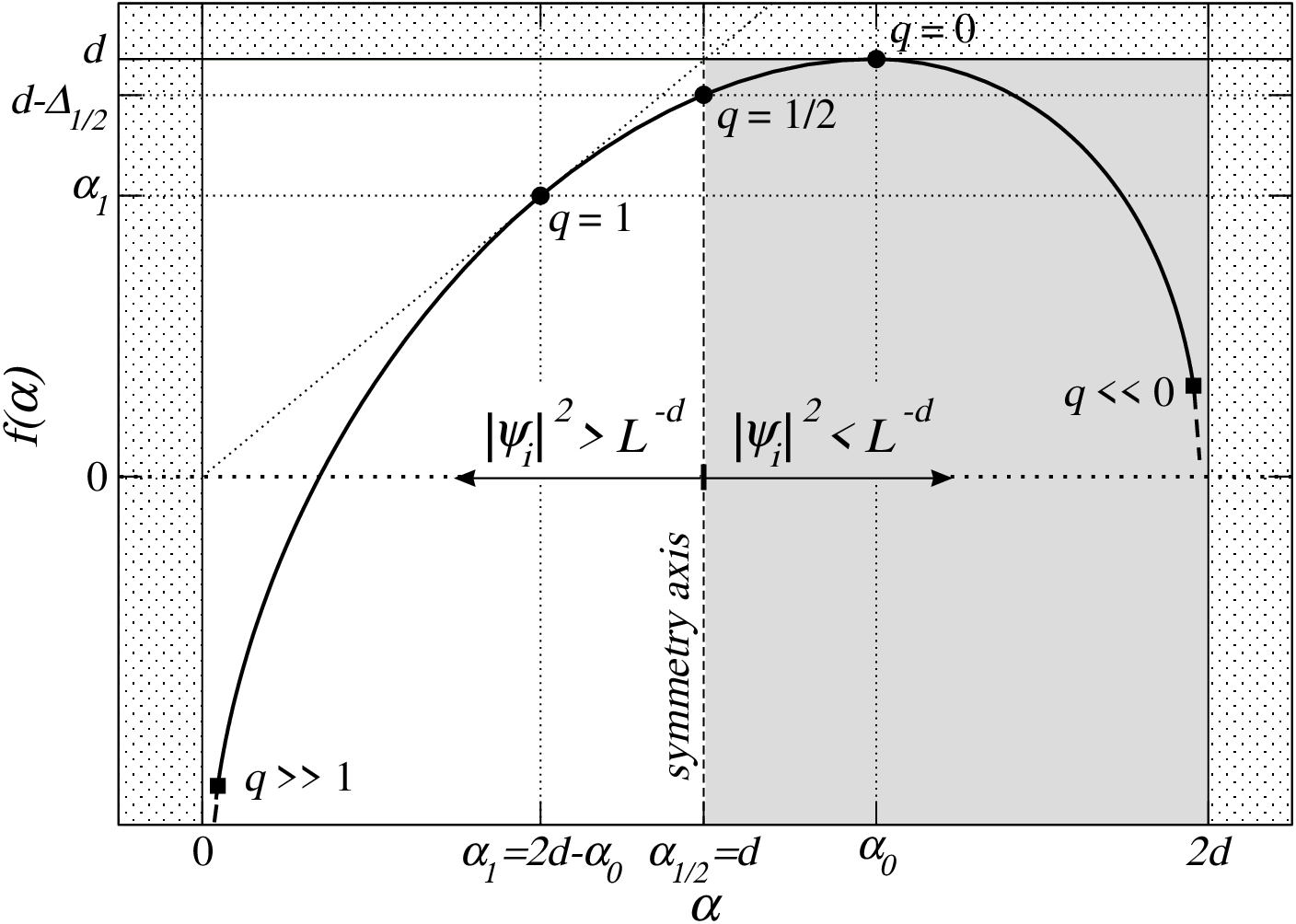}
	\caption{Pictorial representation of the general features of the multifractal spectrum at criticality. The dotted black areas highlight forbidden regions for $f(\alpha)$. The grey shaded area can be connected to the white area through the symmetry relation \eqref{eq-symalfaq},\eqref{eq-symfalfa}, and vice versa. These two regions are determined by different  wavefunction amplitudes, $|\psi_i|^2 > L^{-d}$ (white) and $|\psi_i|^2<L^{-d}$ (light grey). The properties of the spectrum at the points corresponding to $q=1/2$, and the symmetry-related $q=0$, and $q=1$ are explicitly included: $f(\alpha_0)=d$, $f(\alpha_1)=\alpha_1$, $f^\prime(\alpha_1)=1$, $f(\alpha_{1/2}=d)=d-\Delta_{1/2}$.}
	\label{fig-sexyfalfa}
\end{figure}
A pictorial sketch of the expected properties of $f(\alpha)$ is shown in Fig.~\ref{fig-sexyfalfa}.

So far the symmetry has been numerically found in different critical models below 3D\cite{MirFME06,ObuSFGL07}. In a previous work the role of the symmetry law in the MIT for the 3D Anderson model  was thoroughly studied by the authors using the typical average of the scaling law for the gIPR.\cite{VasRR08a} This has been usually regarded as the preferred way to perform MFA. In this work we study the alternative \textit{ensemble average} of the scaling law, and how it performs concerning the symmetry relation. This latter method  manifests itself as the most adequate technique to carry out the numerical MFA, leading to an even better agreement with \eqref{eq-symalfaq}.
\section{MFA using ensemble average}
\label{sec-mfa}
The numerical MFA is based on an \textsl{averaged} form of the scaling law for the gIPR in the limit $\lambda\equiv{l}/{L}\rightarrow 0$, where the contributions from all finite-size critical wavefunctions are propertly taken into account.
The scaling law for the ensemble average involves the arithmetic average of $P_q$ over all realizations of disorder,
\begin{equation}
 	\left<P_q (\lambda) \right>\propto\lambda^{\tau^\textrm{ens}(q)},
	\label{eq-ENSscale}
\end{equation}
where $\langle\cdots\rangle$ denotes the arithmetic average over all states.
Thus the definition of the scaling exponents is
\begin{equation}
 	\tau^\textrm{ens} (q) = \lim_{\lambda\rightarrow 0}\frac{\ln \langle P_q (\lambda)\rangle}{\ln \lambda},
	\label{eq-tauens}
\end{equation}
and the corresponding definitions of $\alpha$ and $f(\alpha)$ can be written in a compact form as
\begin{subequations}
\begin{align}
	\alpha^\textrm{ens}_q & =  \lim_{\lambda\rightarrow0}\;\frac{1}{\ln\lambda}\left\langle \sum_{k=1}^{N_\lambda}\widetilde{\delta}_k(q,\lambda)\ln\widetilde{\delta}_k(1, \lambda)\right\rangle
	\label{eq-alfaens}, \\
	f^\textrm{ens}_q \equiv f(\alpha^\textrm{ens}_q)& =  \lim_{\lambda\rightarrow0}\;\frac{1}{\ln\lambda}\left< \sum_{k=1}^{N_\lambda}\widetilde{\delta}_k(q,\lambda)\ln\widetilde{\delta}_k(q,\lambda)\right>.
	\label{eq-fens}
\end{align}
\label{eq-falfaens}
\end{subequations}
Here $\widetilde{\delta}_k(q,\lambda)\equiv \mu_k^q(\lambda) / \langle P_q(\lambda)\rangle $, which is not normalized for every wavefunction but after the average over all of them. Let us emphasize that although Eqs.~\eqref{eq-falfaens} are handy analytically, it is much more useful for numerical purposes to develop them in longer expressions with simpler factors (see Sec.~\ref{sec-size}).

In contradistinction to the typical average \cite{VasRR08a} which is determined by the behaviour of representative wavefunctions,
the ensemble average weighs the contribution of all wavefunctions equally, including rare
(and hence not representative)  realizations of the disorder. These rare events are indeed responsible for the negative values of $f(\alpha)$. Therefore it is very important to take them into account by doing the ensemble average, if one wants to have a complete picture of the singularity spectrum.
We emphasize that in the thermodynamic limit both averaging processes  must provide the same singularity spectrum in the positive region.
The relation between typical and ensemble averaging has been previously commented in the literature.\cite{MirE00, EveMM08}
\section{Data for the 3D Anderson model at criticality}
\label{sec-model}
The standard tight-binding Anderson model,\cite{BraK03} with uniform diagonal disorder of mean $0$ and width $W_c$ and nearest-neighbour hopping is considered in a cubic lattice of volume $L^3$. The band width is $6$ at zero disorder.
We use the critical disorder at $W_c=16.5$ and for every disorder realization the $L^3\times L^3$ Hamiltonian is diagonalized with periodic boundary conditions to obtain the five eigenstates closest to the band centre $E=0$. \cite{SleMO03} The whole set of data used for the analysis, including system sizes and number of samples for each,  is described in detail in Table \ref{tab-data}. We refer the reader to Ref.~\onlinecite{VasRR08a} for more technical details and to Refs.\ \onlinecite{BraK03} and \onlinecite{Mor04} for recent comprehensive reviews of the subject.
\begin{table}
\caption{Linear system sizes $L$, volume $V$,  number of eigenstates and total wavefunction values $\psi_i$, used for the numerical MFA.}
\begin{tabular}{r r r r}
\hline
\hline
$L$ & $V=L^3$          &  samples&  $\psi_i$ \mbox{    }\\\hline
20  & $8\times 10^3$   &  24 995 &  $2\times 10^8$\\
30  & $9\times 10^3$   &  25 025 &  $6.8\times 10^8$\\
40  & $6.4\times 10^4$ &  25 025 &  $1.6\times 10^9$\\
50  & $1.3\times 10^5$ &  25 030 &  $3.1\times 10^9$\\
60  & $2.2\times 10^5$ &  25 030 &  $5.4\times 10^9$\\
70  & $3.4\times 10^5$ &  24 950 &  $8.6\times 10^9$\\
80  & $5.1\times 10^5$ &  25 003 &  $1.3\times 10^{10}$\\
90  & $7.3\times 10^5$ &  25 005 &  $1.8\times 10^{10}$\\
100 & $1\times 10^6$   &  25 030 &  $2.5\times 10^{10}$\\
140 & $2.7\times 10^6$ &  105    &  $2.9\times 10^8$\\
160 & $4.1\times 10^6$ &  125    &  $5.1\times 10^8$\\
180 & $5.8\times 10^6$ &  100    &  $5.8\times 10^8$\\
200 & $8\times 10^6$   &  100    &  $8\times 10^8$\\
210 & $9.3\times 10^6$ &  105    &  $9.7\times 10^8$\\
240 & $1.4\times 10^7$ &  95     &  $1.3\times 10^9$\\
\hline
\hline
\end{tabular}
\label{tab-data}
\end{table}
\section{Scaling with box size}
\label{sec-box}
The easiest way to approach the thermodynamic limit in the scaling law \eqref{eq-ENSscale}
is considering the limit $l\rightarrow 0$ for the box size $l$. Using this method, we only need realizations for a system with a fixed linear size $L$, that is partitioned equally into an integer number of smaller boxes of linear size $l$.
This is the same partitioning scheme that we have previously considered when studying the typical average of the scaling law. \cite{VasRR08a}

For each state, the $q$-th moment of the box probability $\mu_k^q(l)$ is evaluated in each box, and $P_q$ is obtained by summing the contribution from all boxes.
The scaling behaviour  \eqref{eq-ENSscale} is then obtained for different values of $l$. In all the computations the values of the box size ranges in the interval $10\leqslant l \leqslant L/2$. 
\subsection{General features of $f^\textrm{ens}(\alpha)$}
\label{ssec-BS-genex}
\begin{figure}
  \centering
  \includegraphics[width=\figwidth]{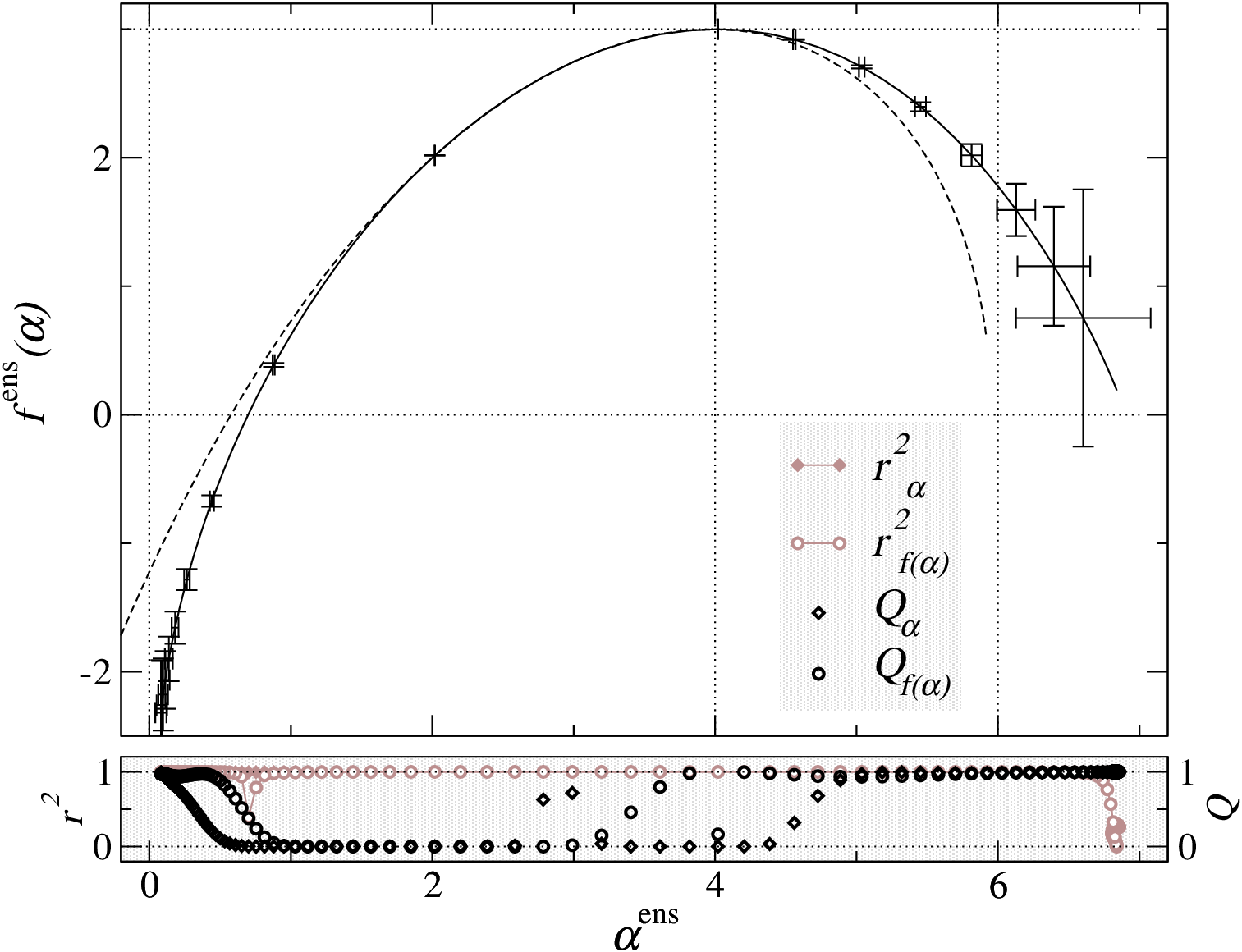}
  \caption{Singularity spectrum (black line) obtained using box-size scaling of the ensemble average of $P_q$ for
system size $L = 100$ with $2.5\times10^4$ states.  The error bars are equal to one standard deviation. The corresponding symmetry-transformed spectrum  $f(2d-\alpha)=f(\alpha)+d-\alpha$ is shown in black dashed line. The values for the linear correlation coefficient $r^2$ and quality-of-fit parameter $Q$ for both $\alpha^\textrm{ens}$ and $f^\textrm{ens}(\alpha)$  are shown in the bottom shaded panel.}
\label{fig-3Dbox-ens-best}
\end{figure}
\begin{figure}
  \centering
  \includegraphics[width=\figwidth]{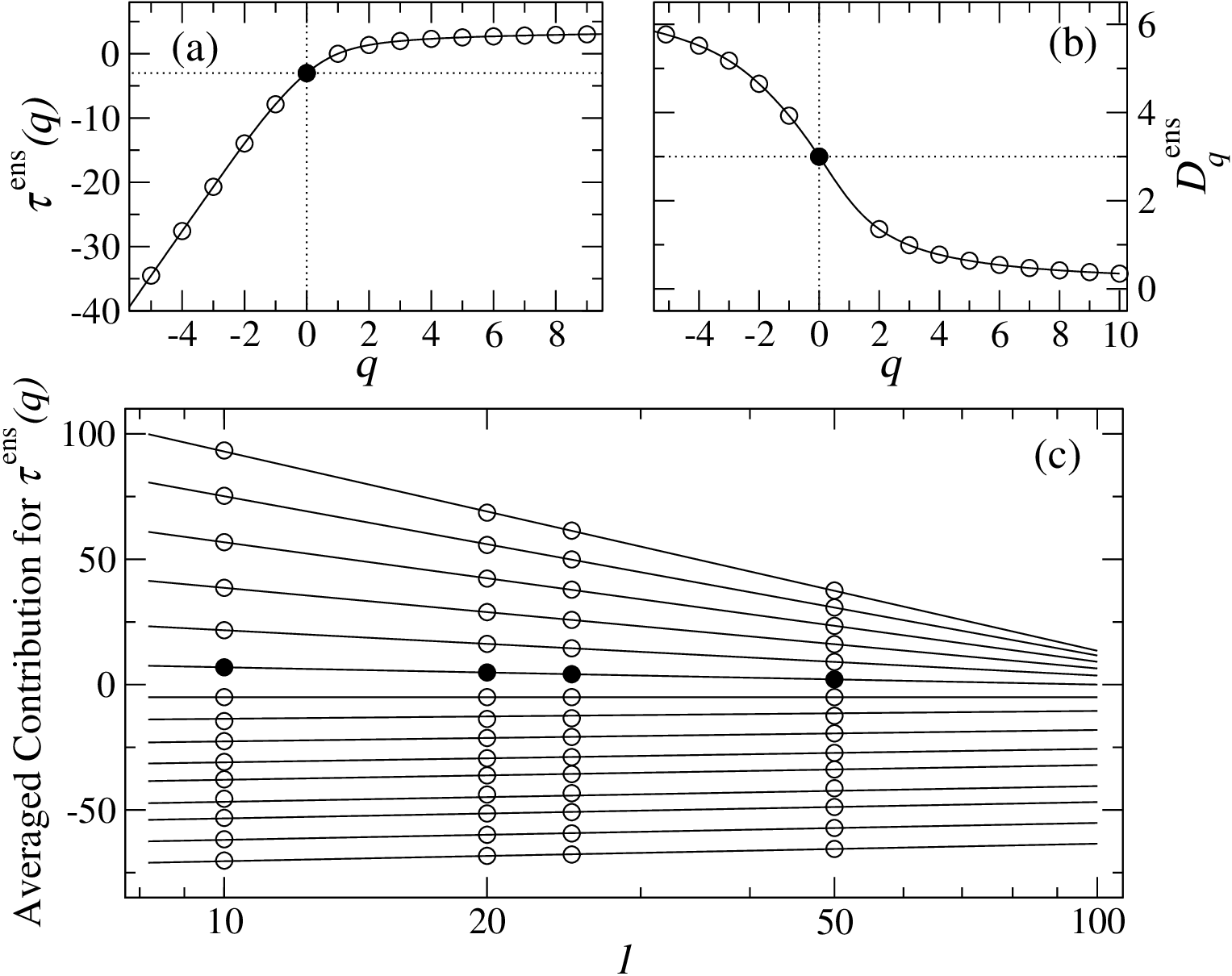}
  \caption{(a) Mass exponents $\tau^\textrm{ens}(q)$  and  (b) generalized fractal dimensions $D_q^\textrm{ens}$ corresponding to the singularity spectrum in Fig.~\ref{fig-3Dbox-ens-best}. Dashed lines in upper panels highlight the values $D_0=d$ and $\tau_0=-d$. Symbols highlight integer values of $q$.
Panel (c):  linear fits of Eq.~\eqref{eq-tauens}. Only fits for integer values of $q$ ranging from $q=-5$ (top) to $q=9$ (bottom) are shown. The value of $\tau^\textrm{ens}(q)$ is given by the slope of the fits. Data points for $q\neq 0$ have been properly shifted vertically  for optimal visualization. Data for $q=0$ highlighted with filled symbols. Standard deviations are contained within symbol size in all panels. }
\label{fig-3Dbox-ens-tau}
\end{figure}
The singularity spectrum for $L=100$ having $2.5\times 10^4$ states is shown in Fig.~\ref{fig-3Dbox-ens-best} with its symmetry transformed spectrum.
The first thing to notice is that the $f^\textrm{ens}(\alpha)$ spectrum attains negative values in the region of small $\alpha$, corresponding to high values of the wave function amplitudes. The negative region of the multifractal spectrum describes the scaling of certain sets of unusual values of $|\psi_i^2|$ which only occur for rare critical functions. Let us recall that $f(\bar{\alpha})<0$ is the fractal dimension of the set of points where $|\psi_i^2|\sim L^{-\bar{\alpha}}$, which implies that the number of such points decreases with the system size as $L^{-|f(\bar{\alpha})|}$. These negative dimensions are then determined by events whose probability of occurrence decreases with the system size. The negative part of the spectrum provides valuable information about the distribution of wavefunction values for a finite-size system near the critical point and is needed to give a complete characterization of the multifractal nature of the critical states at the metal-insulator transition.
At the left-half part of $f^\textrm{ens}(\alpha)$ in Fig.~\ref{fig-3Dbox-ens-best}, we observe its termination in the negative region towards $\alpha\rightarrow 0$. The values of $\alpha$ and $f(\alpha)$ are obtained from the slopes of the linear fit of Eqs.~\eqref{eq-falfaens} via a general $\chi^2$ minimization taking into account the statistical uncertainty of the averaged right-hand side terms. The behaviour of the linear correlation coefficient $r^2$ and the quality-of-fit parameter $Q$ for the different parts of the spectrum (corresponding to different values of the moments $q$) is shown in the bottom panel of Fig.~\ref{fig-3Dbox-ens-best}.  The $r^2$ value is very near to one for almost all $\alpha$ which shows the near perfect linear behaviour of the data points.  The parameter $Q$ gives an estimation on how reliable the fits are according to the error bars of the points involved in the fits. The unusual decrease of $Q$ observed around $\alpha=3$, corresponding to $q\sim 0.5$, in Fig.~\ref{fig-3Dbox-ens-best} is due to an underestimation of the standard deviations of the points in the fits, since the linear correlation coefficient is still very high in this region.
It can also be seen that the uncertainties for the points of $f^\textrm{ens}(\alpha)$ tend to grow when approaching the ends of the spectrum. This effect is more significant when doing ensemble average, but it should be naturally expected since the higher the value of $|q|$ is, the more the numerical inaccuracies of $|\psi_i|^2$ are enhanced, specially in the region of negative $q$, corresponding to the right branch of the spectrum. The mass exponents $\tau^\textrm{ens}(q)$ and the fits of Eq.~\eqref{eq-tauens} are shown in Fig.~\ref{fig-3Dbox-ens-tau}, along with the generalized fractal dimensions $D^\textrm{ens}_q\equiv \tau^\textrm{ens}(q)/(q-1)$ corresponding to the spectrum in Fig.~\ref{fig-3Dbox-ens-best}.
\subsection{Effects of system size and disorder realizations on $f^\textrm{ens}(\alpha)$}
\label{ssec-BS-effects}
\begin{figure}
  \centering
  \includegraphics[width=\figwidth]{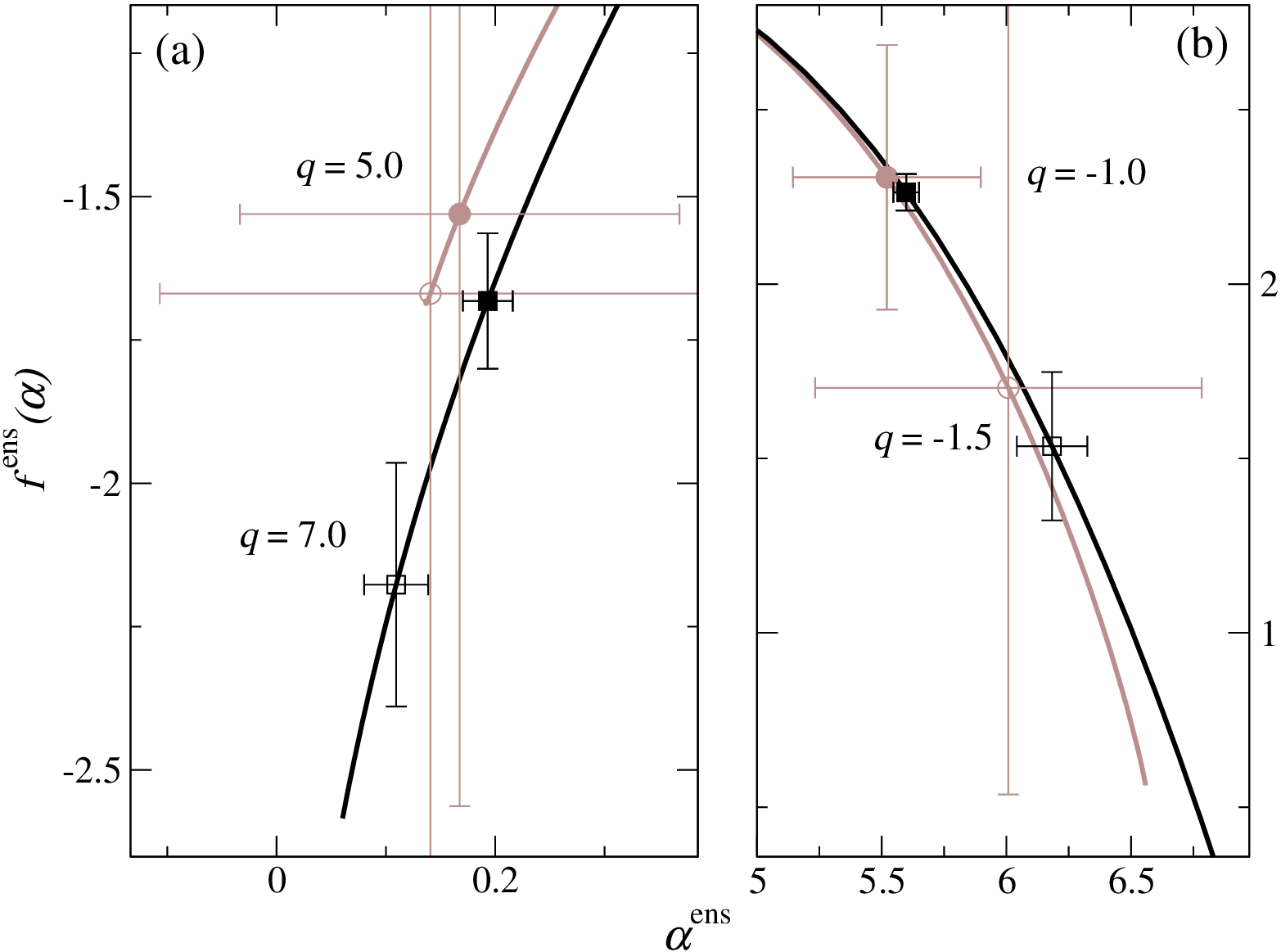}
  \caption{Left (a) and right (b) branches of the singularity spectrum obtained using box-size scaling and ensemble average for
system size $L = 60$ with $2.5\times10^2$ (grey) and $2.5\times10^4$ (black) number of states.
The filled symbols denote $q=5.0$ (a) and $q=-1.0$ (b).  The empty symbols mark $q=7.0$ (a) and $q=-1.5$ (b).
The error bars are equal to one standard deviation.}
\label{fig-3Dbox-ens-noofstates}
\end{figure}
In Fig.~\ref{fig-3Dbox-ens-noofstates} we study the effects of the number of states and disorder realizations on $f^\textrm{ens}(\alpha)$ for $L=60$ having $2.5\times10^2$ and $2.5\times10^4$ states.  Considering two particular $q$ values at each tail, when the number of samples is increased
we see that the domain of $f^\textrm{ens}(\alpha)$ is enlarged. The point corresponding to a given $q$ appears later in the spectrum and thus the left end reaches more negative values with more states [Fig.~\ref{fig-3Dbox-ens-noofstates}(a)]. The same stretching effect can also be observed for the right branch in Fig.~\ref{fig-3Dbox-ens-noofstates}(b). Additionally the reliability of the data points in the singularity spectrum is significantly improved as shown by the huge decrease in their uncertainties. These effects prove the strong dependence of the ensemble averaging on the number of samples taken.

\begin{figure}
  \centering
  \includegraphics[width=\figwidth]{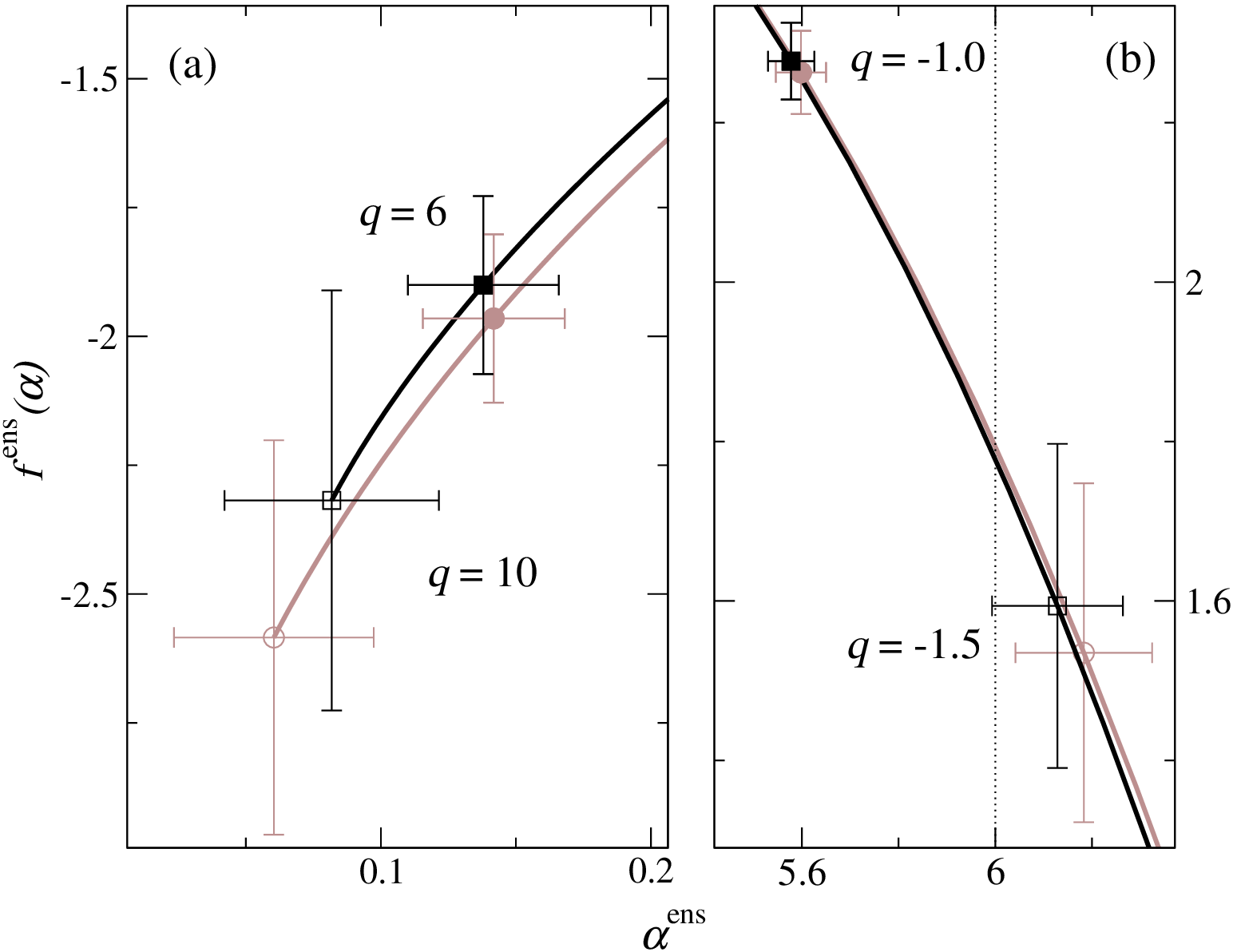}
  \caption{Left (a) and right (b) branches of the singularity spectrum obtained using box-size scaling and ensemble average
for system sizes $L = 60$ (grey)  and $L = 100$ (black) where each has $2.5\times10^4$ states.
The filled symbols denote $q=6.0$ (a) and $q=-1.0$ (b).  The empty symbols mark $q=10.0$ (a) and $q=-1.5$ (b).
The error bars indicate one standard deviation.}
\label{fig-3Dbox-ens-varyL}
\end{figure}
The effect of the system size on the shape of the singularity spectrum is presented in Fig.~\ref{fig-3Dbox-ens-varyL}. Here we consider system sizes $L=60$ and $L=100$ each having $2.5\times10^4$ number of states.  Once again, we take two particular $q$ values at each tail as shown in panels (a) and (b) and observe how their locations change when the system size is varied. When we consider a bigger system size with the same number of realizations, the domain of $f^\textrm{ens}(\alpha)$ tends to decrease, and so for the same $q$ range, we see less negative values at the left end [Fig.~\ref{fig-3Dbox-ens-varyL}(a)]. In other words to be able to observe the same extent of the negative $f^\textrm{ens}(\alpha)$ values of $L=60$, one must average over more states when a bigger system size such as $L=100$ is considered. The same shrinking effect also occurs in the right branch of the spectrum. This unexpected behaviour is due to the nature of the ensemble averaging process, that is strongly determined by the contribution of rare events which are less likely to happen for larger systems. This important effect will be discussed in more detail in Sec.~\ref{sec-size}.
\subsection{Symmetry relation}
\label{ssec-BS-sym}
\begin{figure}
  \centering
  \includegraphics[width=\figwidth]{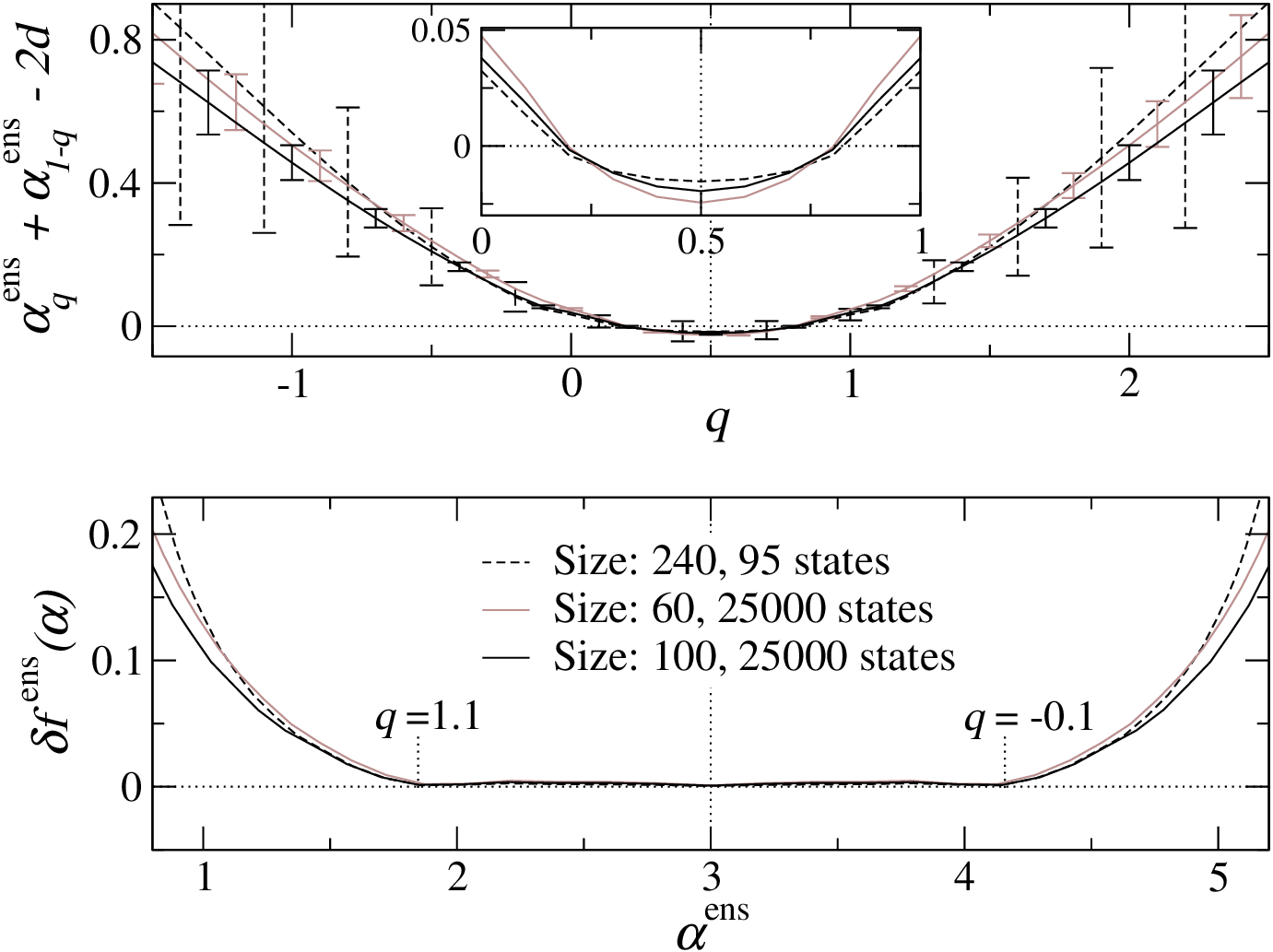}
  \caption{Measure of degree of symmetry of the multifractal spectrum obtained from ensemble average doing scaling with box size. The upper panel shows the  numerical evaluation of the symmetry law as a function of $q$ for system sizes
$L=240$ (dashed black) with $95$ states, $L=60$ (gray) with $2.5\times10^4$ states and
$L=100$ (solid black) for $2.5\times10^4$ states.
The bottom panel shows $\delta f(\alpha)$ versus $\alpha$. There's no correspondence between the abscissa axes of the upper and lower plots. For clarity, two values of $q$ for the black line are explicitly written.}
\label{fig-3Dbox-ens-symdeg}
\end{figure}
In the upper panel of Fig.~\ref{fig-3Dbox-ens-symdeg} we give a numerical evaluation of the symmetry law.
An approximate estimation of the symmetry law is also shown in the lower panel using \eqref{eq-symfalfa}, \cite{VasRR08a} which measures the distance between the spectrum and its symmetry-transformed counterpart.
We compare data for  $L= 240$ ($95$ states), $L=60$ ($2.5\times10^4$ states) and $L=100$ ($2.5\times10^4$ states). Our results show that in general the
closest agreement to the symmetry in the singularity spectrum is achieved for the cases with the highest number of disorder realizations, in particular for $L=100$ ($f(\alpha)$ shown in Fig.~\ref{fig-3Dbox-ens-best}). Although around the symmetry point $q=1/2$ the spectrum obtained using the largest system size available, $L=240$ with $95$ states, tends to behave slightly better (inset in upper panel of Fig.~\ref{fig-3Dbox-ens-best}), the tendency is inverted when looking at a broader region of $q$ values.
This result is a clear manifestation of how important the number of disorder realizations is when doing ensemble average.

It is therefore clear that the more realizations, the better the symmetry is. Obviously a bigger system size helps reduce finite-size effects, but we have shown that
increasing system size also implies generating more states in order to obtain the same extent of $f^\textrm{ens}(\alpha)$.
Thus from a numerical viewpoint an agreement between system size and disorder realizations must be found to optimize the use of box-size scaling and ensemble averaging.
\section{Scaling with system size}
\label{sec-size}
The scaling with the system size may be the most adequate way to describe the thermodynamic limit of the scaling law for the gIPR
\eqref{eq-ENSscale} ($L\rightarrow\infty$), however the numerical eigenstate problem is highly demanding for very large 3D systems. \cite{SchBR06}

The formulae \eqref{eq-tauens} and \eqref{eq-falfaens} for the singularity spectrum  are now affected by the substitution:
$\lim_{\lambda\rightarrow 0}\Rightarrow -\lim_ {L\rightarrow\infty}$. As for the typical average, \cite{VasRR08a} the box size $l$ which determines the integrated probability distribution is set to $l=1$ for non-negative moments ($q\geqslant 0$) and to a value $l>1$ (usually $l=5$) for $q<0$, in order to minimize the errors and the uncertainties in the right branch of $f(\alpha)$.
For the case $l=1$ the formulae to obtain the spectrum reduce to
\begin{subequations}
\begin{align}
	- \alpha^\textrm{ens}_q \ln L &\sim \frac{\left<\sum_i |\psi_i|^{2q} \ln |\psi_i|^2\right>}{\left<\sum_j |\psi_j|^{2q}\right>}, \\
	- f^\textrm{ens}_q \ln L &\sim \frac{\left<\sum_i |\psi_i|^{2q} \ln |\psi_i|^{2q}\right>}{\left<\sum_j |\psi_j|^{2q}\right>}
	-  \ln \left<\sum_i |\psi_i|^{2q}\right>.
\end{align}
\label{eq-falfa-ens-psi}
\end{subequations}
We note  the clear difference between the ensemble average \eqref{eq-falfa-ens-psi} and the typical average techniques [Eqs.~(14) in Ref.~\onlinecite{VasRR08a}]. The values of $\alpha^\textrm{ens}_q$ and $f^\textrm{ens}_q$ are obtained from the slopes of the linear fits of the averaged terms in Eqs.~\eqref{eq-falfaens} ($q\geqslant 0$) and \eqref{eq-falfa-ens-psi} ($q<0$) versus $\ln L$, for different values of the system size $L$.
\subsection{General features of $f^\textrm{ens}(\alpha)$}
\label{ssec-SS-genex}
\begin{figure}
  \centering
  \includegraphics[width=\figwidth]{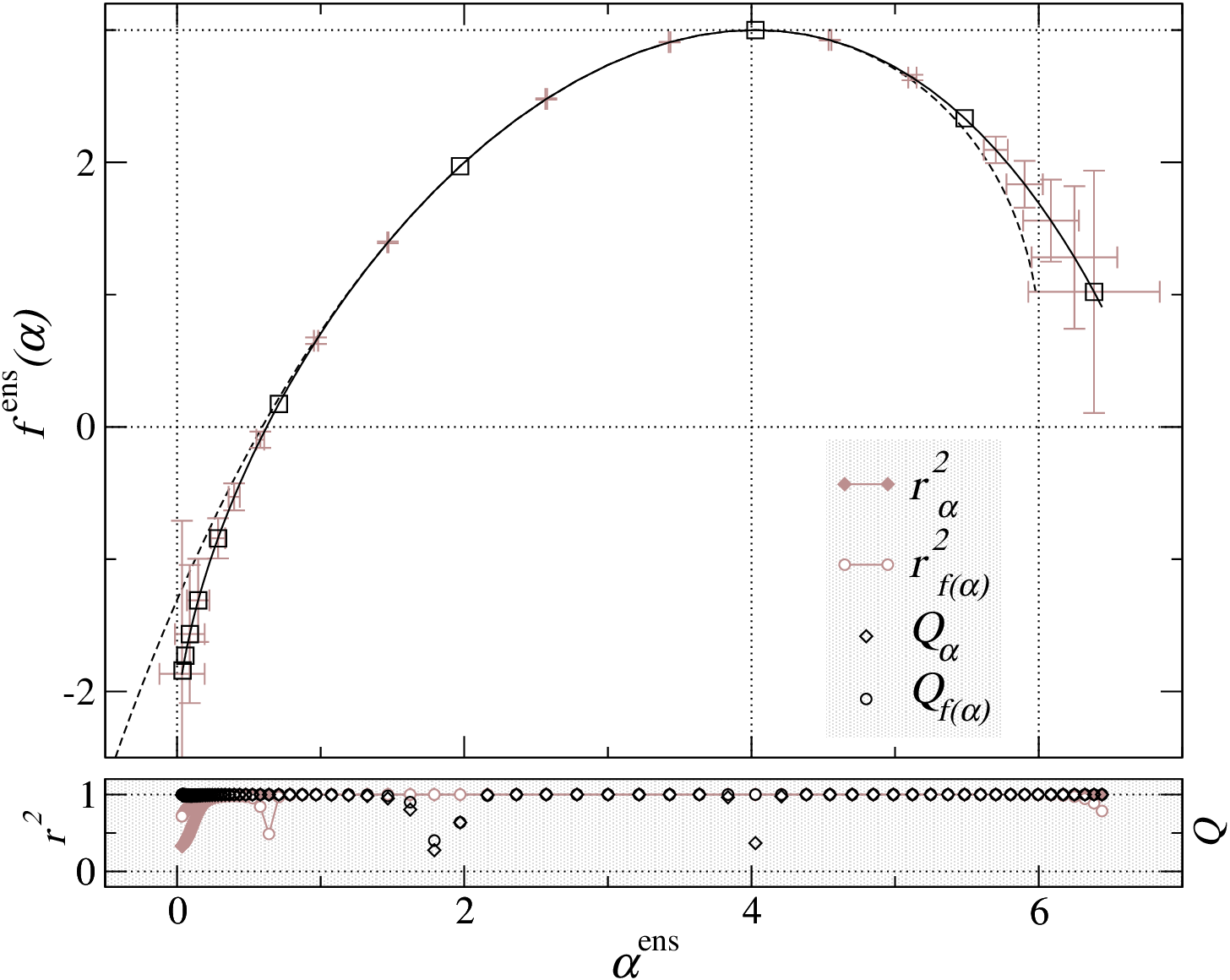}
  \caption{Singularity spectrum obtained from ensemble averaging. System sizes from $L=20$ to $100$ have been considered with $\sim2.5\times10^4$ different
wavefunctions for each system size as shown in Table \ref{tab-data}. The dashed line is the symmetry-transformed spectrum. The values of $q$ range from $q=-2$ to $q=7$ with a step of $0.1$ ($l=1$ for $q \geqslant 0$ and $l=5$ for $q<0$). Symbols highlight the values corresponding to integer $q$. Error bars in grey are standard deviations. The lower panel shows the linear correlation coefficient ($r^2$) and the quality-of-fit parameter ($Q$) of the linear fits to obtain the values for $\alpha$ and $f(\alpha)$.}
\label{fig-3Dsize-ens}
\end{figure}

\begin{figure}
  \centering
  \includegraphics[width=\figwidth]{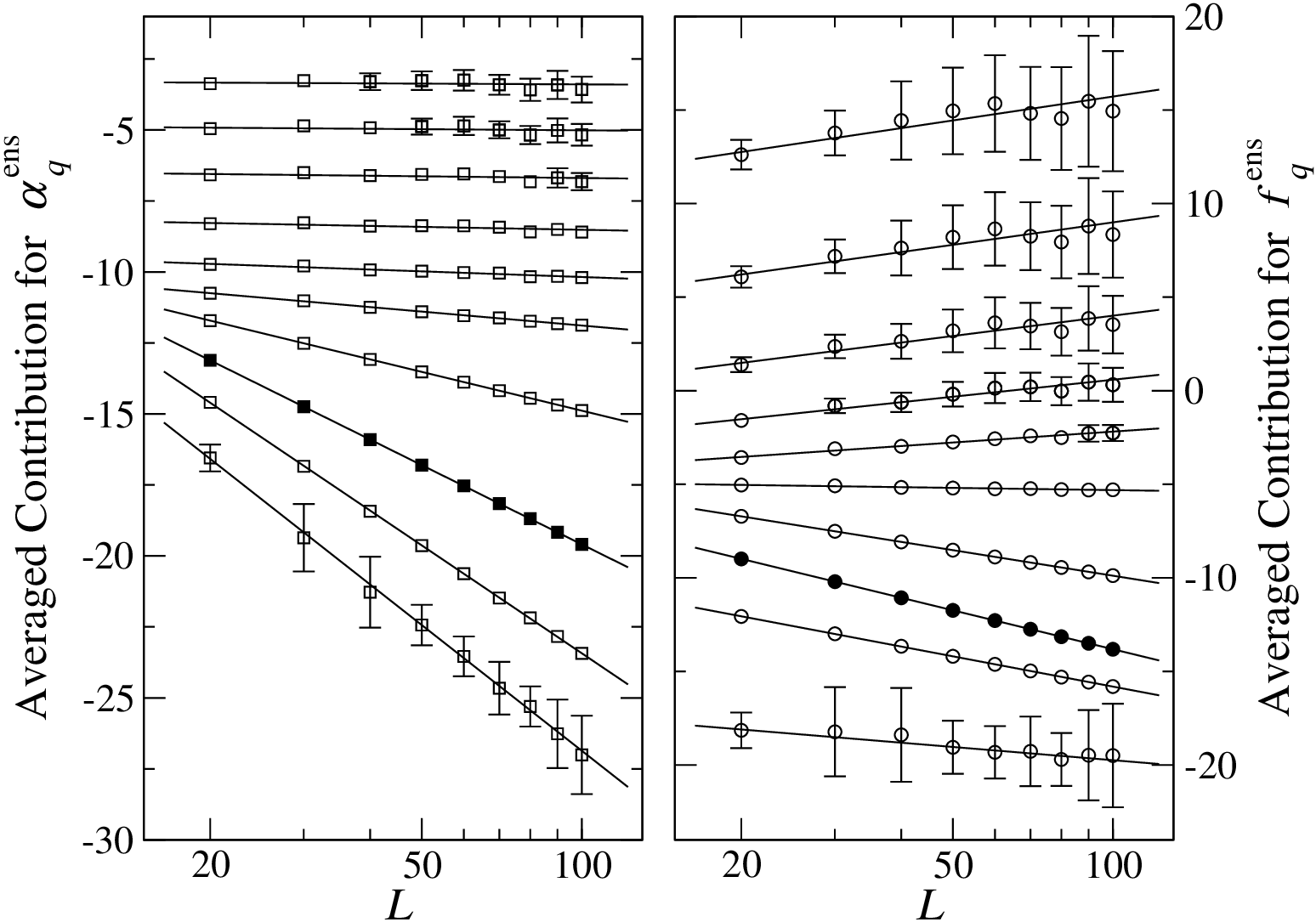}
  \caption{Linear fits of Eqs.~\eqref{eq-falfaens} for  $\alpha^\textrm{ens}_q$ values (left) and $f^\textrm{ens}_q$ values (right) of the singularity spectrum in Fig.~\ref{fig-3Dsize-ens}. Only fits for integers values of $q$ ranging from $q=7$ (top) to $q=-2$ (bottom) are shown. The values of $\alpha^\textrm{ens}_q$ and
$f^\textrm{ens}_q$ are given by the slopes of the fits. Data points for $q\neq 0$ have been properly shifted vertically to ensure optimal visualization.  Data for $q=0$ highlighted with filled symbols. When not shown, standard deviations are contained within symbol size.}
\label{fig-3Dsize-ens-fits}
\end{figure}

The multifractal spectrum obtained from the ensemble average is shown in Fig.~\ref{fig-3Dsize-ens}, where we have considered $9$ different linear system sizes ranging from $L=20$ to $100$ for the scaling after averaging over $\sim2.5\times10^4$ states for each size as shown in Table \ref{tab-data}.
The branch of negative values characterizing $f^\textrm{ens}(\alpha)$ can be clearly seen.
The absence of an infinite slope in the spectrum when crossing the abscissa axis must also be emphasized. As discussed in Ref.~\onlinecite{VasRR08a} this  confirms the fact that the divergence of the slope at the termination points observed when doing the typical average, $f^\textrm{typ}(\alpha)$, is purely a finite-size effect, since both averages must provide the same result for $f(\alpha)\geqslant 0$ in the thermodynamic limit. This is also supported by the systematic shift of the left end of $f^\textrm{typ}(\alpha)$ to smaller values of $\alpha$ whenever more states or larger system sizes are considered. \cite{VasRR08a}
The error bars for the values of $f^\textrm{ens}(\alpha)$ are considerably larger than the ones obtained for $f^\textrm{typ}(\alpha)$ using the same system sizes and disorder realizations [Fig.~(7) in Ref.~\onlinecite{VasRR08a}]. This is of course a consequence of having larger errors for the points used in the linear fits, shown in Fig.~\ref{fig-3Dsize-ens-fits}.
These higher uncertainties are due to the nature of the average itself and the probability distribution function for the generalized IPR. The probability density for $P_q$ is an asymmetric function around its maximum with long tails,\cite{MirE00,EveM00} resembling a log-normal distribution. The calculation of the arithmetic average of $P_q$, involved in the ensemble average is therefore much more heavily based on the number of disorder realizations than the determination of the geometric mean used for the typical average, and  thus larger uncertainties and slower convergence must be expected for the ensemble-averaged situation with the same number of wavefunctions.
Regarding the errors in the values of $f^\textrm{ens}(\alpha)$, it is remarkable how their magnitude grows, for high values of $|q|$, apparently at the same rate as the spectrum deviates from the symmetry-transformed curve (dashed line in Fig.~\ref{fig-3Dsize-ens}). This suggests that it might be possible to observe almost a perfect agreement with the symmetry law, using this small range of system sizes for the scaling, if the number of realizations were large enough.
%
\subsection{Effects of the number of disorder realizations on $f^\textrm{ens}(\alpha)$}
\label{ssec-SS-states}
\begin{figure}
  \centering
  \includegraphics[width=\figwidth]{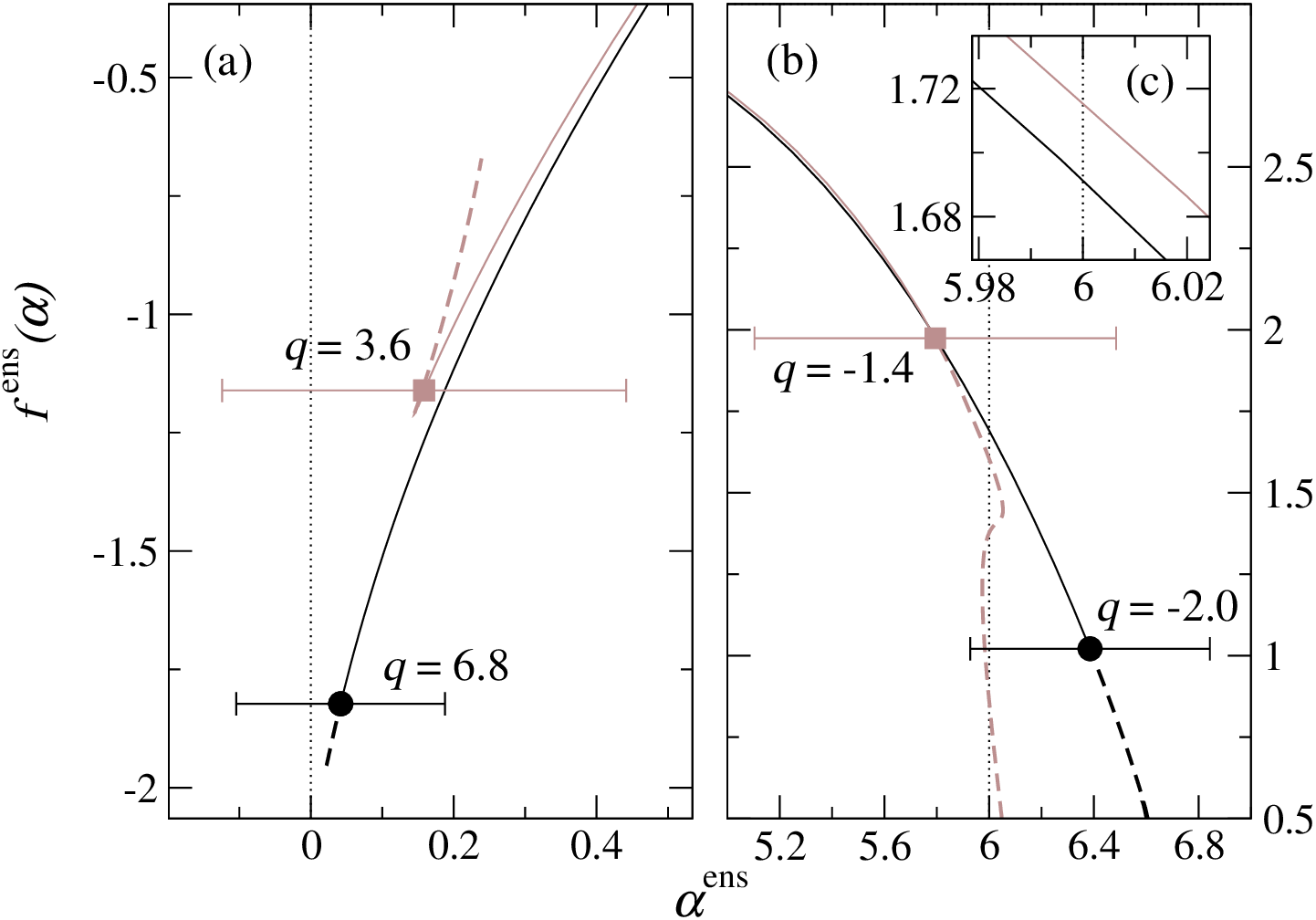}
  \caption{Left (a) and right(b) branches of the singularity spectrum obtained from ensemble averaging scaling with system sizes from $L=20$ to $100$ and same  number of states for each size: (grey) $10^3$, (black) $\sim2.5\times10^4$. The values of  $q$ range from $q=-10$ to $q=10$ with a step of $0.1$ ($l=1$ for $q\geqslant 0$ and $l=5$ for $l<0$). The vertical standard deviation for the points marked with filled symbols is always $\sigma_{f}=1.0$, and only the uncertainty for $\alpha$ has been included, for clarity. The $q$ value corresponding to each of the symbols is indicated. Dashed lines represent the spectrum in each case for higher values of $q$. Inset (c) shows the change in the value $f(\alpha=6)$ for the cases: (grey) $5\times10^3$ states for each size, (black) $\sim2.5\times10^4$ states for each size.}
\label{fig-3Dsize-ens-stateseffect}
\end{figure}
The effect of increasing the number of states in the ensemble average can be seen in Fig.~\ref{fig-3Dsize-ens-stateseffect}, for scaling with $L\in [20, 100]$.  A reduction of the standard deviations must be expected whenever more realizations are taken into account. To make this clear we have considered two situations: averaging over $10^3$ states or over $\sim2.5\times 10^4$ states for each size.
In Fig.~\ref{fig-3Dsize-ens-stateseffect} the points with the specified vertical uncertainty appear later (for higher values of $q$) on the left and right branches of the spectrum when we increase the number of states in the average. This clearly means that for a fixed position on the $f^\textrm{ens}(\alpha)$ curve, the uncertainty shrinks when more states are included. There is however, another significant effect that must be emphasized. In Fig.~\ref{fig-3Dsize-ens-stateseffect}(a)(b), the spectrum obtained for values of $q$ higher than the ones indicated is represented by dashed lines. For the average including only $10^3$ samples for each size, the values of the spectrum for high $|q|$ are completely absurd and $f(\alpha)$ behaves in an unexpected way,  showing ``kinks'' and bumps as a consequence of a loss of precision in the fits caused by very large uncertainties. This implies that by increasing the number of states in the ensemble average not only the standard deviations are reduced for each point, but the domain of accessible values for $f(\alpha)$ is also enlarged, e.g. for more wavefunctions the spectrum reaches more negative values [Fig.~\ref{fig-3Dsize-ens-stateseffect}(a)].  This is in marked contrast to the typical average \cite{VasRR08a} for which one obtains almost the same range of $f(\alpha)$ independently of the number of states considered. When doing the ensemble average, the appearance of these ``kinks'' in the spectrum, either for system-size or box-size scaling where they have also been observed,  is always the fingerprint of a lack of sampling of the distributions, i.e. not enough disorder realizations.

In the inset (c) of Fig.~\ref{fig-3Dsize-ens-stateseffect}, we have also illustrated the behaviour of the value $f(\alpha=6)$, at the upper boundary required by the symmetry relation, when the number of states is changed from $5\times10^3$ to $\sim2.5\times10^4$ for each system size. The spectrum tends to be in better agreement with the upper boundary required by \eqref{eq-symfalfa} when the number of disorder realizations increases.
\subsection{Effects of the range of system sizes on $f^\textrm{ens}(\alpha)$}
\label{ssec-SS-sizes}
\begin{figure}
  \centering
  \includegraphics[width=\figwidth]{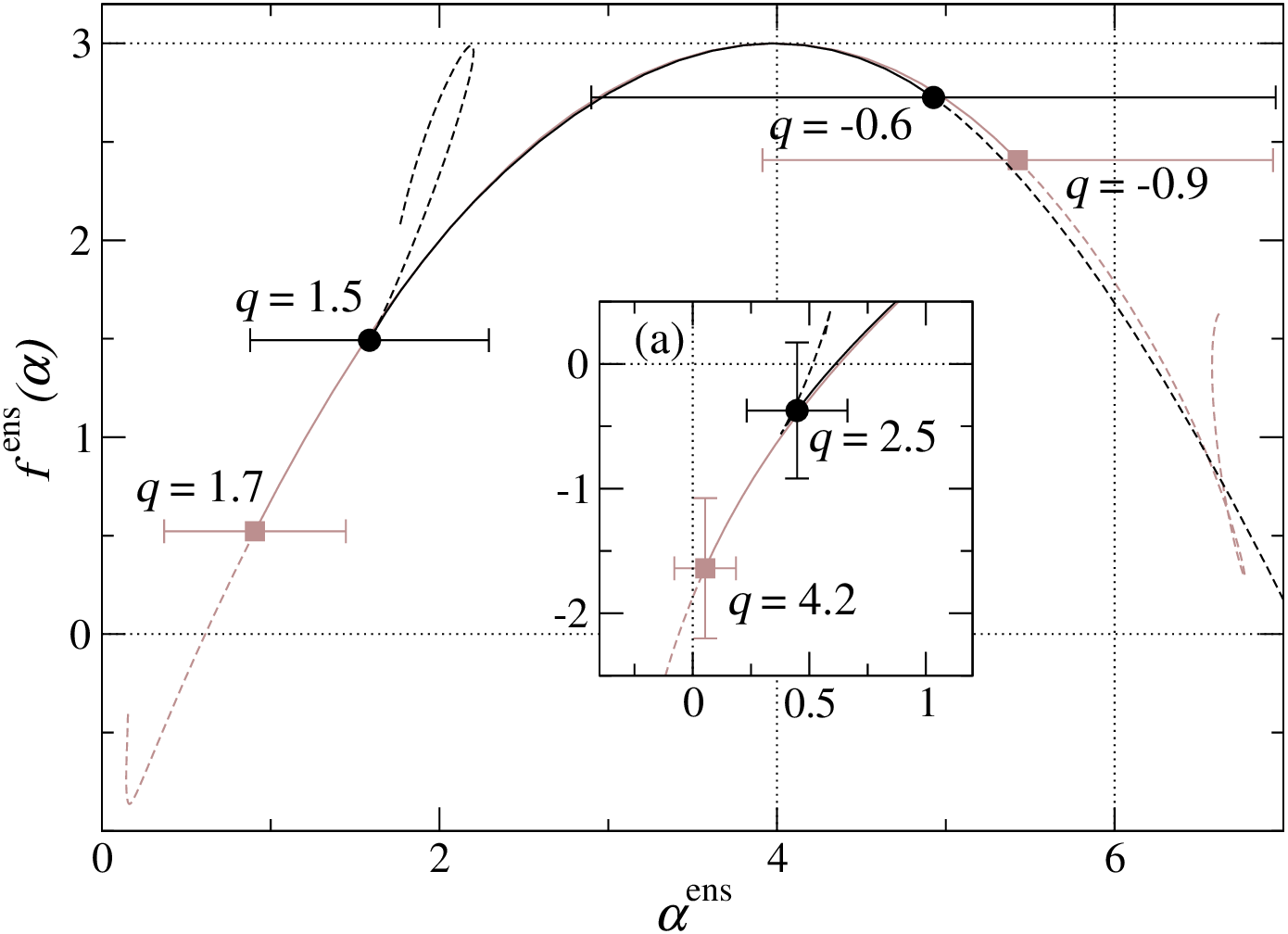}
  \caption{Singularity spectrum obtained from ensemble averaging scaling with $7$ system sizes $L\in [40, 100]$ and $10^2$ states for each size (grey), and $6$ system sizes  $L\in[140,240]$ and $\sim10^2$ states for each size (black). The values of  $q$ range from $q=-10$ to $q=10$ with a step of $0.1$ ($l=1$ for $q\geqslant 0$ and $l=5$ for $l<0$). Filled symbols correspond to points with the same vertical standard deviation ($\sigma_{f}\simeq1.0$, not shown for clarity). Dashed lines represent the spectrum in each case for values of $q$ higher than the ones indicated beside their corresponding points. Inset (a) shows the spectrum for a different set of data: $L\in [20, 60]$ (grey) and $L\in[60,100]$ (black) with  $\sim2.5\times10^4$ states for each size in both cases.}
\label{fig-3Dsize-ens-lengtheffect}
\end{figure}
In order to study the effects of the system size, we show in Fig.~\ref{fig-3Dsize-ens-lengtheffect} the singularity spectrum obtained doing scaling in different intervals: $L\in[40,100]$ and $L\in[140,240]$, taking $\sim100$ wave functions for each size in both cases. The fact that we have only averaged over $100$ states for each size, makes the standard deviations noticeably large, however this does not affect the conclusions qualitatively. When we consider larger system sizes for a fixed number of disorder realizations, the region where we can reliably obtain the multifractal spectrum shrinks. Moreover if we go to high enough values of $|q|$ (highlighted by dashed lines in Fig.~\ref{fig-3Dsize-ens-lengtheffect}), it can be noticed how the  wrong behaviour of $f(\alpha)$ is enhanced. This is a very counterintuitive result, since one would expect that for increasing system sizes, the number of disorder realizations needed to obtain the spectrum  with a given degree of reliability should decrease proportionally -- that is in fact what happens with the typical averaging. \cite{VasRR08a} However for ensemble averaging the conclusion is just the opposite: if you want to improve the spectrum in a given region of the tails and you consider larger system sizes to reduce finite size effects, the number of disorder realizations must also be increased. This is due again to the nature of the ensemble averaging process and the shape of the distributions for the gIPR. The arithmetic average is heavily based on rare events which are less likely to appear for bigger systems, and so the number of realizations has to grow with the system size in order to include the proper amount of rare events.
This can be more clearly understood in the region of negative fractal dimensions. We know that the number of points in a single wave function such that $|\psi_i^2|\sim L^{-\bar{\alpha}}$ where $f(\bar{\alpha})<0$, is $L^{-|f(\bar{\alpha})|}\ll 1$. Therefore to be able to see the region of negative fractal dimensions we would need a number of states $\mathcal{N}$ such that we can guarantee $\mathcal{N} L^{-|f(\bar{\alpha})|}\gg 1$. This implies that the number of disorder realizations must go as $\mathcal{N}\sim L^{|f(\bar{\alpha})|}$, and thus it increases with the system size. This effect can be observed in the inset (a) of Fig.~\ref{fig-3Dsize-ens-lengtheffect}, where we have compared scaling with sizes $L\in [20, 60]$ and $L\in[60,100]$ with  $\sim2.5\times10^4$ states for each size in both cases. For higher sizes and the same number of states, we are not able to see the same region of negative fractal dimensions. Aside from this effect, it must nevertheless be emphasized that when we consider larger system sizes, the right branch of the spectrum tends to find a better agreement with the upper boundary required by the symmetry law.
\subsection{Symmetry relation}
\label{ssec-SS-sym}
\begin{figure}
  \centering
  \includegraphics[width=\figwidth]{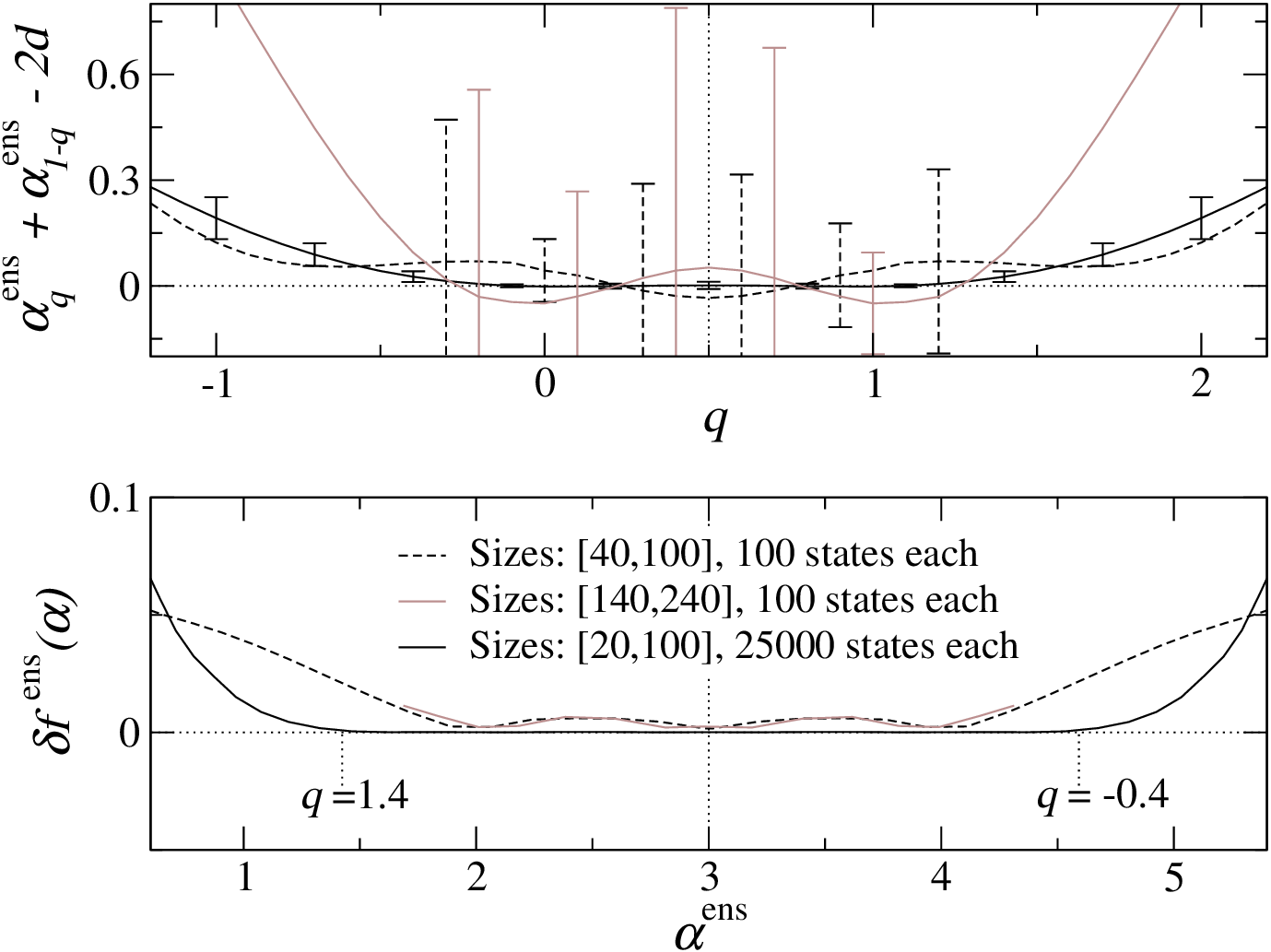}
  \caption{Measure of degree of symmetry of the multifractal spectrum obtained from ensemble average doing scaling with system size. The upper panel shows the  numerical evaluation of the symmetry law as a function of $q$. Dashed black: $7$ system sizes from $L=40$ to $100$ and $10^2$ states for each. Grey: $6$ system size from $L=140$ to $240$ and $\sim10^2$ states for each. Solid black: $9$ system sizes from $L=20$ to $100$ and $\sim2.5\times 10^4$ states for each. The bottom panel shows $\delta f(\alpha)$ versus $\alpha$. There's no correspondence between the abscissa axes of the upper and lower plots. For clarity, two values of $q$ for the black line are explicitly written.}
\label{fig-3Dsize-ens-symdeg}
\end{figure}
To discuss the fulfilment of the symmetry using ensemble average and scaling with system size, let us look at Fig.~\ref{fig-3Dsize-ens-symdeg} where the numerical
evaluation of the symmetry law \eqref{eq-symalfaq} is shown for different ranges of system sizes and disorder realizations. The best result, according
to the symmetry, corresponds undoubtedly to the case with the highest number of disorder realizations, $\sim2.5\times 10^4$, for which the scaling analysis involves sizes from $L=20$ to $100$. The difference is remarkable between the situation corresponding to (i) $L\in[40,100]$ averaging over $100$ states only, where the symmetry is hardly satisfied at all, and (ii) the best case where the development of the plateau for $\alpha_q+\alpha_{1-q} -2d$ around $q=0.5$ can be seen very clearly. The spectrum obtained for $L\in[120,240]$ with $\sim100$ states for each size also deviates noticeably from the symmetry. These differences can also be seen in the bottom panel of Fig.~\ref{fig-3Dsize-ens-symdeg} where the degree of symmetry is estimated by $\delta f(\alpha)$.

For the ensemble average going to very large system sizes is not the best strategy unless one can generate an increasing number of states. It must be clear that of
course finite size effects will be reduced using large sizes but the number of wavefunctions used for the average has to grow with the system size considered. For a given range of system sizes, increasing the number of states improves the reliability of data, enlarges the accessible domain of $f(\alpha)$, specially in the region of negative dimensions, and improves the symmetry.
\section{Comparison of different scaling and averaging approaches}
\label{sec-typVSens}
\begin{figure}
  \centering
  \includegraphics[width=\figwidth]{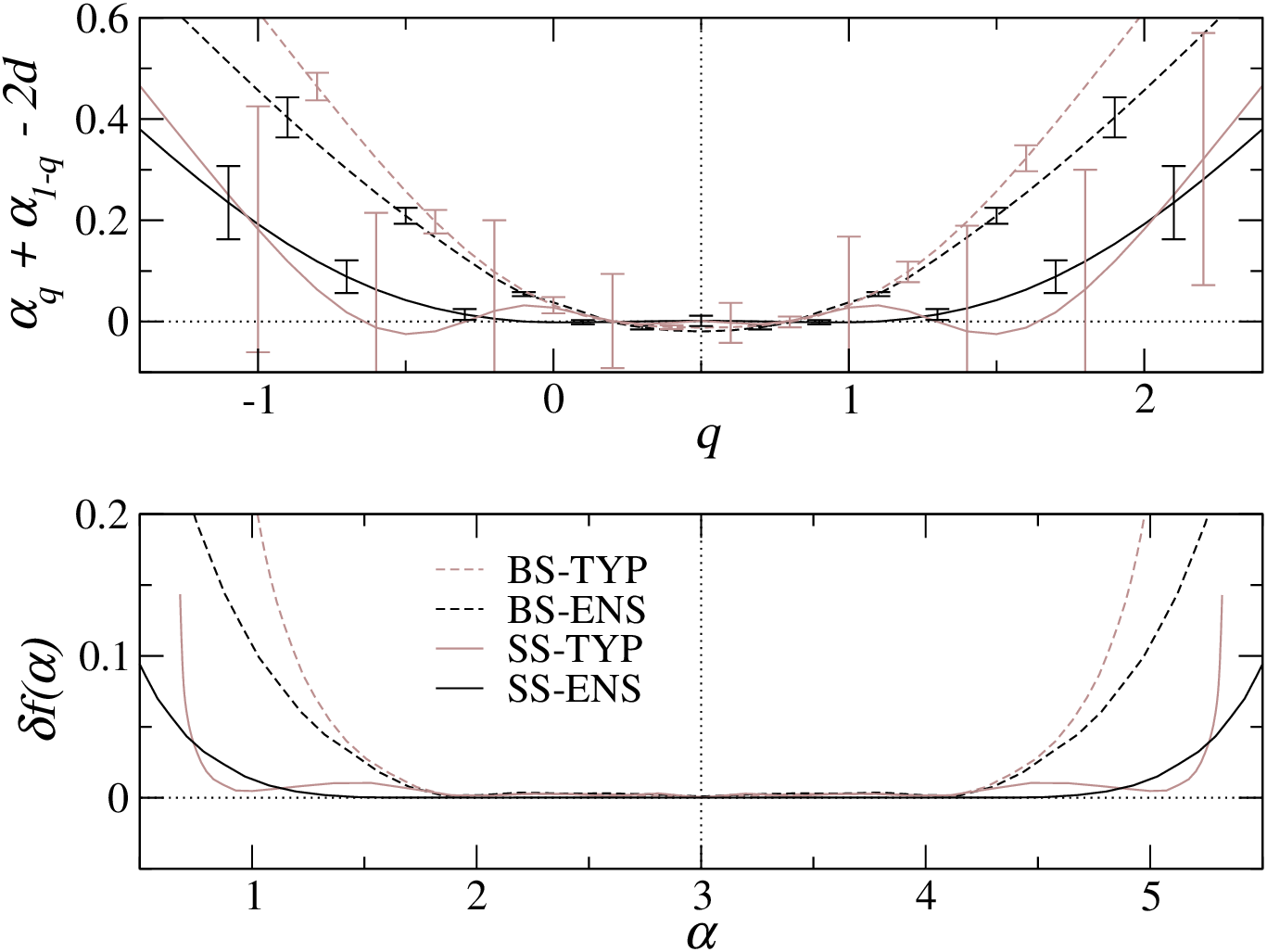}
  \caption{Comparison of degree of symmetry for spectra obtained using: box-size scaling typical average [BS-TYP] (dashed grey),
box-size scaling ensemble average [BS-ENS] (dashed black), system-size scaling typical average [SS-TYP] (solid grey), system-size scaling ensemble average [SS-ENS] (solid black). The best spectrum for each case has been considered: [BS-TYP] $L=240$ ($95$ states), [BS-ENS], $L=100$ ($2.5\times 10^4$ states), [SS-TYP] $L\in[140,240]$  ($\sim10^2$ states for each size), [SS-ENS] $L\in[20,100]$ ($\sim2.5\times10^4$ states for each size). Data for typical average extracted from Ref.~\onlinecite{VasRR08a}. The upper panel shows the  numerical evaluation of the symmetry law as a function of $q$. The bottom panel shows $\delta f(\alpha)$ versus $\alpha$. There's no correspondence between the abscissa axes of the upper and lower plots.}
\label{fig-symdeg-finalcomparison}
\end{figure}

 Taking the symmetry relation \eqref{eq-symalfaq} as a measure of the quality of the numerical MFA, let us compare the results of the different scaling and averaging techniques. In Fig.~\ref{fig-symdeg-finalcomparison} we show the best analyses obtained from box-size scaling and system-size scaling using typical average and ensemble average in both cases. Data corresponding to the typical average has been extracted from Ref.~\onlinecite{VasRR08a}.
The performance of the system-size scaling technique (solid lines) is clearly much better than box-size scaling (dashed lines). This is not a very surprising result, since one expects finite-size effects to be more enhanced in box-size scaling. For each of the scaling procedures the ensemble average (black) is also better than the typical average (grey). This may be not be so intuitive, since due to the nature of the distribution functions for $P_q$, \cite{MirE00,EveM00} one might expect the typical average to be a better choice. However it turns out that the ensemble average does better in revealing the true behaviour in the thermodynamic limit.

Let us  recall the strategies that give the best result for each of the techniques.
For typical average the best symmetry is achieved using the largest system sizes available, either for box-size or system-size scaling, although the number of realizations is not the highest. \cite{VasRR08a} On the other hand, using ensemble average, the safest choice is to consider smaller system sizes for which a very large number of disorder realizations can be obtained.
\section{Conclusions and outlook}
\label{sec-concl}
In this work we have studied the symmetry law \eqref{eq-symalfaq} for the multifractal spectrum of the electronic states
at the metal-insulator transition in the 3D Anderson model, using the ensemble-averaged scaling law of the gIPR \eqref{eq-ENSscale}.
A detailed analysis has revealed how the MFA is affected by system size and number of samples.
System-size scaling with ensemble average has manifested itself as the most adequate method to perform numerical MFA, in contrast to box-size scaling and typical average which had been mainly the method of choice in previous studies. \cite{MilRS97,GruS95,SchG91}  Since the ensemble average is strongly based on the number of disorder realizations, from a numerical point of view, the best strategy to carry out the analysis is to consider a sensible range of system sizes for which a very large number of states can be generated.

All our results suggest that the symmetry law is true in the thermodynamic limit, since a better agreement is found whenever a high enough number of disorder realizations and larger system sizes are considered. The symmetry relation \eqref{eq-symfalfa} then provides a powerful tool to obtain a complete picture of $f(\alpha)$ at criticality, since it would suffice to obtain numerically the spectrum in the most reliable region, $q>0$, and apply the symmetry to complete the function for $q<0$.

\begin{figure}
  \centering
  \includegraphics[width=\figwidth]{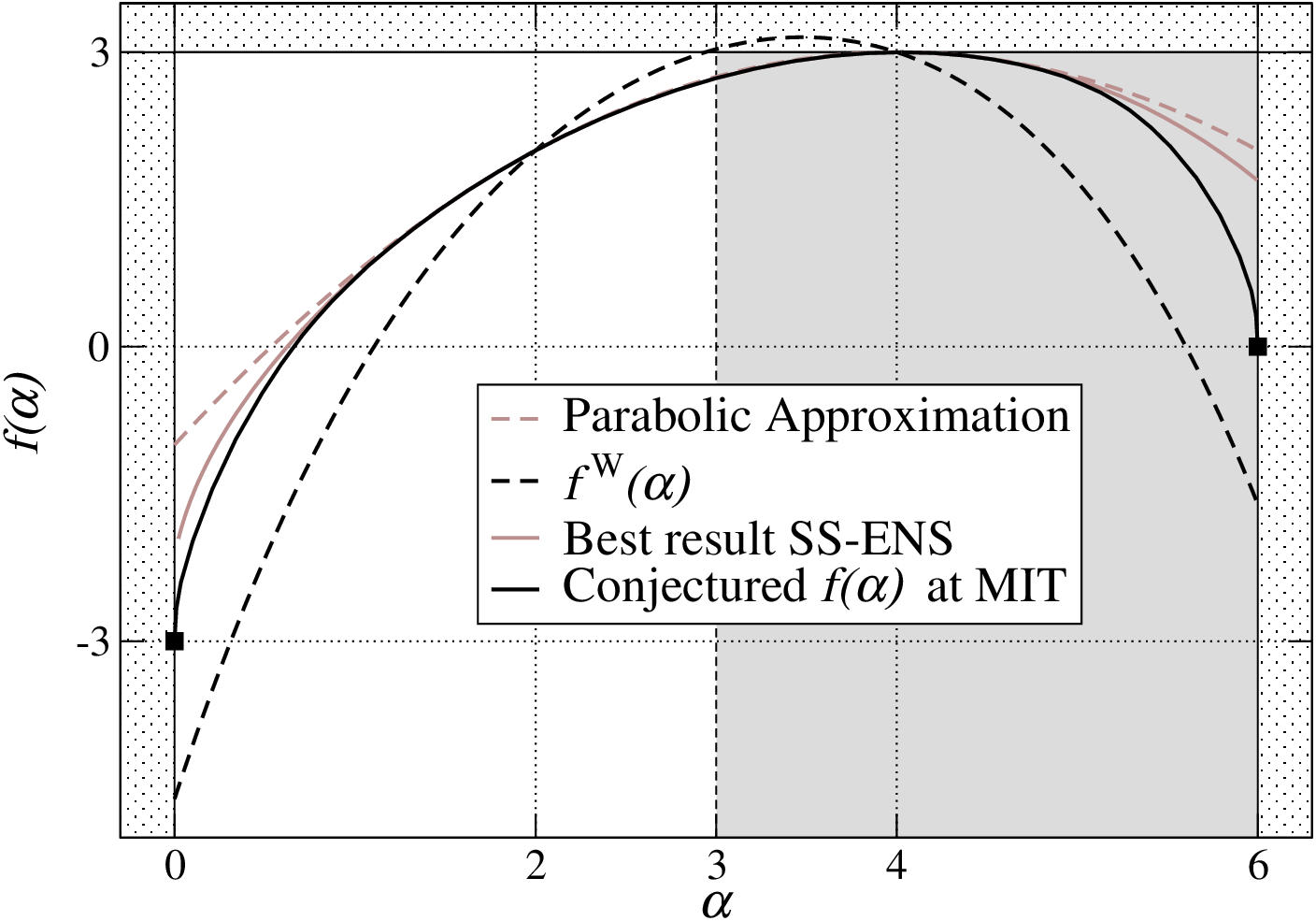}
  \caption{Singularity spectra given by the Parabolic Approximation (dashed gray), Eq.~\eqref{eq-Wegner} (dashed black), numerical calculation doing scaling with system-size and ensemble average described in Fig.~\ref{fig-3Dsize-ens} (solid gray) and conjectured $f(\alpha)$ at the MIT for the 3D Anderson model (solid black). The dotted black areas highlight forbidden regions for $f(\alpha)$. The grey shaded and white areas can be connected through the symmetry relation \eqref{eq-symalfaq},\eqref{eq-symfalfa}.}
\label{fig-conjectured}
\end{figure}

The results obtained for $f(\alpha)$ also provide some useful information about the validity of previous analytical results.
The perturbative analysis in $d=2+\epsilon$ made by Wegner \cite{Weg89} led to the following spectrum
\begin{equation}
\begin{split}
	f^\textrm{W}(\alpha) =& d -\frac{[\alpha-(d+\epsilon)^2]}{4\epsilon} \\ &- \frac{\zeta (3)}{64}{[(\alpha-d)^2-\epsilon^2][(\alpha-d)^2+15\epsilon^2]}
\end{split}
\label{eq-Wegner}
\end{equation}
where $\zeta(x)$ denotes the Riemann zeta function. The first two terms in \eqref{eq-Wegner} constitute the usual parabolic approximation (PA). The extra quartic term is an overestimation in 3D ($\epsilon=1$), as explicitly stated by Wegner, which gives a non-acceptable spectrum as can be seen in  Fig.~\ref{fig-conjectured}. To obtain the correct $f(\alpha$) at $\epsilon=1$ all the other terms in the perturbation series are required.
Remarkably Wegner's result  obeys the symmetry relation \eqref{eq-symfalfa}, provided the spectrum is indeed terminated at $\alpha=0$ and $\alpha=2d$. The symmetry can also be easily checked by calculating $\Delta_q$, which can be written as a power series in $\epsilon$ whose coefficients are all invariant under the transformation $q\rightarrow (1-q)$. The results from the numerical MFA seem to suggest that $f(\alpha)$ tends to approach a finite negative value when $\alpha\rightarrow 0$. This would imply that the spectrum has a termination point. The existence of a termination point at $\alpha=0$ requires that the function $\tau(q)$ has a zero slope for $q$ greater than a certain critical value $q_c$. \cite{EveMM08} This latter fact is also suggested in Fig.~\ref{fig-3Dbox-ens-tau}(a).
Moreover the numerically obtained spectrum is between the PA and $f^\textrm{W}(\alpha)$ as shown in Fig.~\ref{fig-conjectured}. In view of this we are tempted to conjecture that perhaps  the PA and $f^\textrm{W}(\alpha)$ are bounds for the spectrum in the thermodynamic limit. In this were true then $f(\alpha)$ must necessarily terminate at finite values. Furthermore let us speculate that the left termination point is $f(\alpha=0)= -d$ which implies $f(\alpha=2d)=0$ according to the symmetry [Fig.~\ref{fig-conjectured}], and thus the spectrum would not attain negative values at its right end, determined by the smallest wavefunction amplitudes. The resulting $f(\alpha)$ at the 3D Anderson transition would hence be close to the black line shown in Fig.~\ref{fig-conjectured}.

In conclusion, in spite of the large amount of information that the numerical analyses, here and in Ref.\ \onlinecite{VasRR08a}, and the symmetry relation \eqref{eq-symalfaq} have provided, the complete picture of the multifractal spectrum for the MIT in 3D is still elusive. In particular further research is needed to confirm the existence of termination points and whether these happen at negative values on both sides, since this has important implications upon the distribution functions of the wavefunction amplitudes near the localisation-delocalisation transition.
\acknowledgments
RAR gratefully acknowledges EPSRC (EP/C007042/1) for financial support. AR acknowledges financial support from the Spanish government under contracts
JC2007-00303, FIS2006-00716 and MMA-A106/2007, and JCyL under contract SA052A07.

%
%

\end{document}